\documentclass[acmtog,nonacm]{acmart}

\acmSubmissionID{492}
\setcopyright{rightsretained}
\acmJournal{TOG}
\acmYear{2025} \acmVolume{44} \acmNumber{4} \acmArticle{} \acmMonth{8} \acmPrice{}\acmDOI{10.1145/3731161}

\ccsdesc{Computing methodologies~Computer graphics, Machine Learning}

\usepackage[utf8]{inputenc} %
\usepackage[T1]{fontenc}    %
\usepackage{url}            %
\usepackage{booktabs}       %
\usepackage{amsfonts}       %
\usepackage{nicefrac}       %
\usepackage{microtype}      %
\usepackage{xcolor}         %
\usepackage{subcaption}
\usepackage[font=small]{caption}
\usepackage{float}
\usepackage{lscape}
\usepackage[lined,ruled,linesnumbered]{algorithm2e}
\usepackage{multirow}
\usepackage{paralist}
\usepackage{enumitem}
\usepackage{bm}
\usepackage{bbm}
\usepackage{mathtools}
\usepackage{amsmath}
\usepackage{dsfont}
\usepackage{color}
\usepackage{colortbl}
\usepackage{siunitx}
\usepackage{comment}
\usepackage{graphicx}
\usepackage{listings}
\usepackage{xspace}
\usepackage{adjustbox}
\usepackage{layouts}
\usepackage{wrapfig}
\usepackage{tabularx}
\usepackage{booktabs}
\usepackage{makecell}

\usepackage{xr}

\PassOptionsToPackage{table}{xcolor}

\renewcommand{\arraystretch}{1.3}

\def\eg{e.g.,\ }               %
\def\vs{vs.\ }                 %
\def\INGP{INGP\ }                %

\definecolor{LavenderBlue}{rgb}{0.7020,    0.8039,    0.8902}
\definecolor{Lightapricot}{rgb}{0.9961,    0.8510,    0.6510}
\definecolor{thirdtablecolor}{rgb}{0.8706,    0.7961,    0.8941}

\definecolor{codegreen}{rgb}{0,0.6,0}
\definecolor{codegray}{rgb}{0.5,0.5,0.5}
\definecolor{codepurple}{rgb}{0.58,0,0.82}
\definecolor{backcolour}{rgb}{0.95,0.95,0.92}
\definecolor{derekblue}{RGB}{144,187,195}
\definecolor{darkderekblue}{RGB}{86,130,140}
\definecolor{verylightgray}{RGB}{245,245,245}
\definecolor{veryverylightgray}{RGB}{253,253,253}
\definecolor{myred}{rgb}{1.0, 0.6, 0.5}
\definecolor{mygreen}{rgb}{0.39, 0.74, 0.21}
\definecolor{myyellow}{rgb}{1.0, 0.9, 0.4}
\definecolor{myblue}{rgb}{0.11,0.29,0.54}
\definecolor{darkgreen}{rgb}{0.13, 0.55, 0.13}

\lstdefinestyle{my_code_style}{
  backgroundcolor=\color{backcolour},
  commentstyle=\color{codegreen},
  keywordstyle=\color{magenta},
  numberstyle=\tiny\color{codegray},
  stringstyle=\color{codepurple},
  basicstyle=\ttfamily\footnotesize,
  breakatwhitespace=false,
  breaklines=true,
  captionpos=b,
  keepspaces=true,
  numbers=left,
  numbersep=5pt,
  showspaces=false,
  showstringspaces=false,
  showtabs=false,
  tabsize=2
}
\lstdefinelanguage{Pseudocode}{
    morekeywords={abstract,break,case,catch,const,continue,do,else,elseif,%
      end,export,false,for,function,immutable,import,importall,if,in,%
      macro,module,otherwise,quote,return,switch,true,try,type,typealias,%
      using,while,either,or,max,abs,foreach,def},%
   sensitive=true,%
   alsoother={\$},
   morecomment=[l]\#,%
   morecomment=[n]{\#=}{=\#}
}[keywords,comments,strings]%

\SetAlFnt{\small}
\SetAlCapFnt{\small}
\SetAlCapNameFnt{\small}
\SetAlCapHSkip{0pt}

\makeatletter
\newcommand{\layoutdetails}{%
\begin{tabular}{ll}
 \texttt{\textbackslash{textwidth}} & \printinunitsof{in}\prntlen{\textwidth} \\
\texttt{\textbackslash{linewidth}} & \printinunitsof{in}\prntlen{\linewidth} \\
Main text font &  \f@size pt \f@family \\
\sffamily \small Caption text font &  \sffamily \small \f@size pt \f@family \\
\end{tabular}%
}
\makeatother

\newcommand{\vx}{\bm{x}}
\newcommand{\vy}{\bm{y}}
\newcommand{\vb}{\bm{b}}

\long\def\ignorethis#1{}

\def\papertitle{Stochastic Preconditioning for Neural Field Optimization}
\newcolumntype{b}{>{\small}X}
\newcolumntype{s}{>{\small\hsize=.4\hsize\centering\arraybackslash}c}
\newcolumntype{z}{>{\small\hsize=.3\hsize\centering\arraybackslash}X}
\newcolumntype{v}{>{\small\hsize=.2\hsize\centering\arraybackslash}X}

\author{Selena Ling}
\affiliation{
  \institution{University of Toronto}
  \country{Canada}
  \ and
  \institution{NVIDIA}
  \country{Canada}
}
\orcid{0000-0001-6458-4488}
\email{selena.ling@mail.utoronto.ca}

\author{Merlin Nimier-David}
\affiliation{
  \institution{NVIDIA}
  \country{Switzerland}
}
\orcid{0000-0002-6234-3143}
\email{mnimierdavid@nvidia.com}

\author{Alec Jacobson}
\affiliation{
  \institution{University of Toronto and Adobe Research}
  \country{Canada}
}
\orcid{0000-0003-4603-7143}
\email{jacobson@cs.toronto.edu}

\author{Nicholas Sharp}
\affiliation{
  \institution{NVIDIA}
  \country{USA}
}
\orcid{0000-0002-2130-3735}
\email{nsharp@nvidia.com}

\begin{document}

\title[\papertitle]{\papertitle}
\begin{teaserfigure}
  \centering
  \vspace{-12.0mm}
  \includegraphics[width=\textwidth]{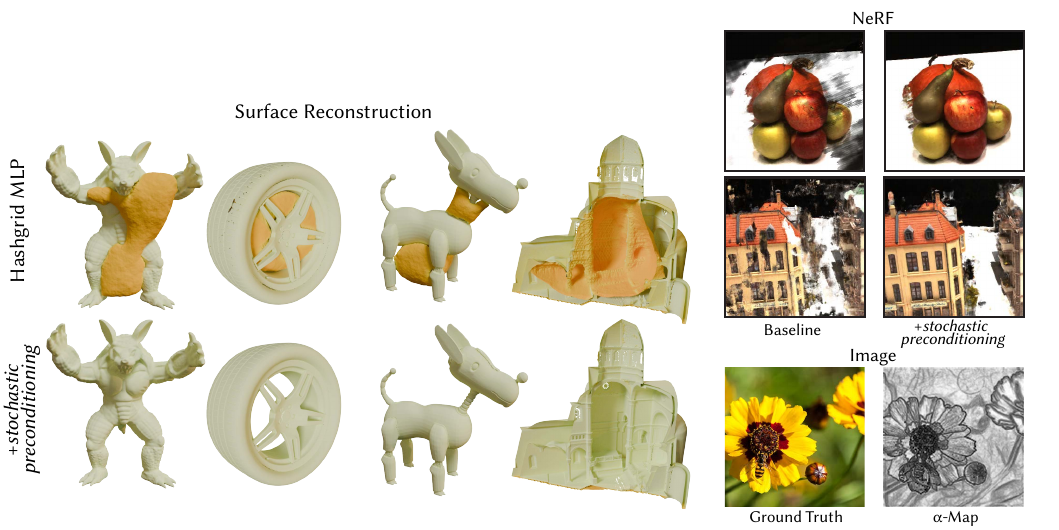}
  \vspace{-5.0mm}
  \caption{
    Stochastic preconditioning adds spatial noise to query locations during neural field optimization; it can be formalized as a stochastic estimate for a blur operator, yet is easily implemented for any field representation in a few lines of code.
    We find that this simple technique eases optimization and significantly improves quality for neural fields optimization tasks like reconstructing signed distance fields from point clouds (\emph{Left}), and neural radiance fields (\emph{Right}), matching or outperforming custom-designed policies and coarse-to-fine schemes.
    The level of noise can even be treated as an optimization variable, naturally yielding a frequency scale maps for the field (\emph{Right, Bottom}).
    \textbf{Webpage:
    \href{https://research.nvidia.com/labs/toronto-ai/stochastic-preconditioning/}{research.nvidia.com/labs/toronto-ai/stochastic-preconditioning/}} }
  \label{fig:teaser}
\end{teaserfigure}

\begin{abstract}
Neural fields are a highly effective representation across visual computing.
This work observes that fitting these fields is greatly improved by incorporating spatial stochasticity during training, and that this simple technique can replace or even outperform custom-designed hierarchies and frequency-space constructions.
The approach is formalized as implicitly operating on a blurred version of the field, evaluated in-expectation by sampling with Gaussian-distributed offsets.
Querying the blurred field during optimization greatly improves convergence and robustness, akin to the role of preconditioners in numerical linear algebra.
This implicit, sampling-based perspective fits naturally into the neural field paradigm, comes at no additional cost, and is extremely simple to implement.
We describe the basic theory of this technique, including details such as handling boundary conditions, and extending to a spatially-varying blur.
Experiments demonstrate this approach on representations including coordinate MLPs, neural hashgrids, triplanes, and more, across tasks including surface reconstruction and radiance fields.
In settings where custom-designed hierarchies have already been developed, stochastic preconditioning nearly matches or improves their performance with a simple and unified approach; in settings without existing hierarchies it provides an immediate boost to quality and robustness.
\end{abstract}

\maketitle

\section{Introduction}
Neural fields represent signals with a function mapping coordinates to signal values, promising an easy-to-optimize and spatially-adaptive representation for signals such as signed distance functions, densities, and radiance fields in 3D, or even simply the color field of an image in 2D.
Many variants have been explored, but the most basic neural field is a function parameterized as a multilayer perceptron (MLP) with weights which are fit to encode the field.
Neural fields can allocate their capacity adaptively as-needed across the domain and benefit from the repertoire of techniques common in machine learning, including auto-differentiation, black-box optimizers, and efficient parallel evaluation on GPUs.

However, neural fields are not without challenges.
Fitting a neural field amounts to solving an expensive nonlinear optimization problem to train the neural network.
This is true even when encoding an already-known signal, and is worsened when optimizing for indirect inverse problems, such as reconstruction from images.
Poor local minima in the optimization landscape lead to low-quality solutions, artifacts, and unresolved detail in the desired signal---in 3D scenes, one way these minima manifest is as so-called floater artifacts, as shown in Figures~\ref{fig:pn-sdf-comparison} and \ref{fig:param-sweep}.

The challenge of optimizing neural fields has led to a proliferation of variants in the representation, with alternate input encodings, architectures, hierarchies, and hybrid strategies, including the well-known work of \citet{nerf, fourierfeat, instantngp}, among many others.
One perspective on these techniques is that they facilitate easier fitting of the field by improving the optimization landscape with a more suitable parameterization for the neural field, or improvements in computational efficiency and expressive power.
However, improved representational power may even worsen the effects of local minima in the optimization process, especially in indirectly supervised and inverse problems with challenging objectives~\cite{diffcd, freenerf, neuralangelo}.
More fundamentally, these many representational variants add significant complexity, and are generally incompatible with one-another.
The challenges of neural fields have also been addressed at the application level, with specialized coarse-to-fine optimization schemes, regularization terms and many other task-specific optimization techniques ~\cite{diffcd, freenerf, neuralangelo, sape, spatialhash, camp, fastscalinginit}.
Though effective, these solutions are often tightly-coupled to the particular task and representation, and do not easily generalize to other neural field problems.

\begin{figure}[b]
  \centering
  \includegraphics[width=\linewidth]{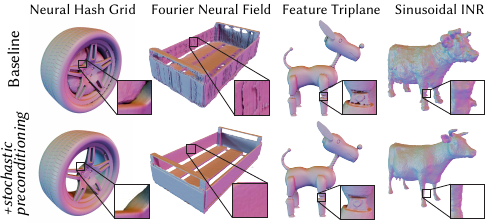}
  \vspace{-2em}
  \caption{
    Stochastic preconditioning can easily be applied to a wide variety of existing neural field representations, consistently reducing artifacts and improving quality, here in the context of fitting signed-distance functions to oriented point clouds.
  }
  \label{fig:pn-sdf-comparison-encodings}
\end{figure}

In this paper, we propose a simple neural field optimization technique that greatly alleviates the challenge of spurious local minima, improves quality, and can be easily integrated into a wide variety of representations and tasks.
The basic idea is to stochastically perturb the input coordinates whenever querying the field during optimization.
From a principled perspective, this is equivalent in-expectation to sampling from a spatial blurring of the field, yet this relationship is implicit and the approach has only the negligible computational cost of adding a small offset to each sample.
We call this technique \textit{stochastic preconditioning} (SP) in reference to the use of preconditioners in linear algebra, to emphasize its benefits in improving the optimization landscape and its applicability to many problems.

\begin{figure*}[t]
    \centering
    \includegraphics[width=\linewidth]{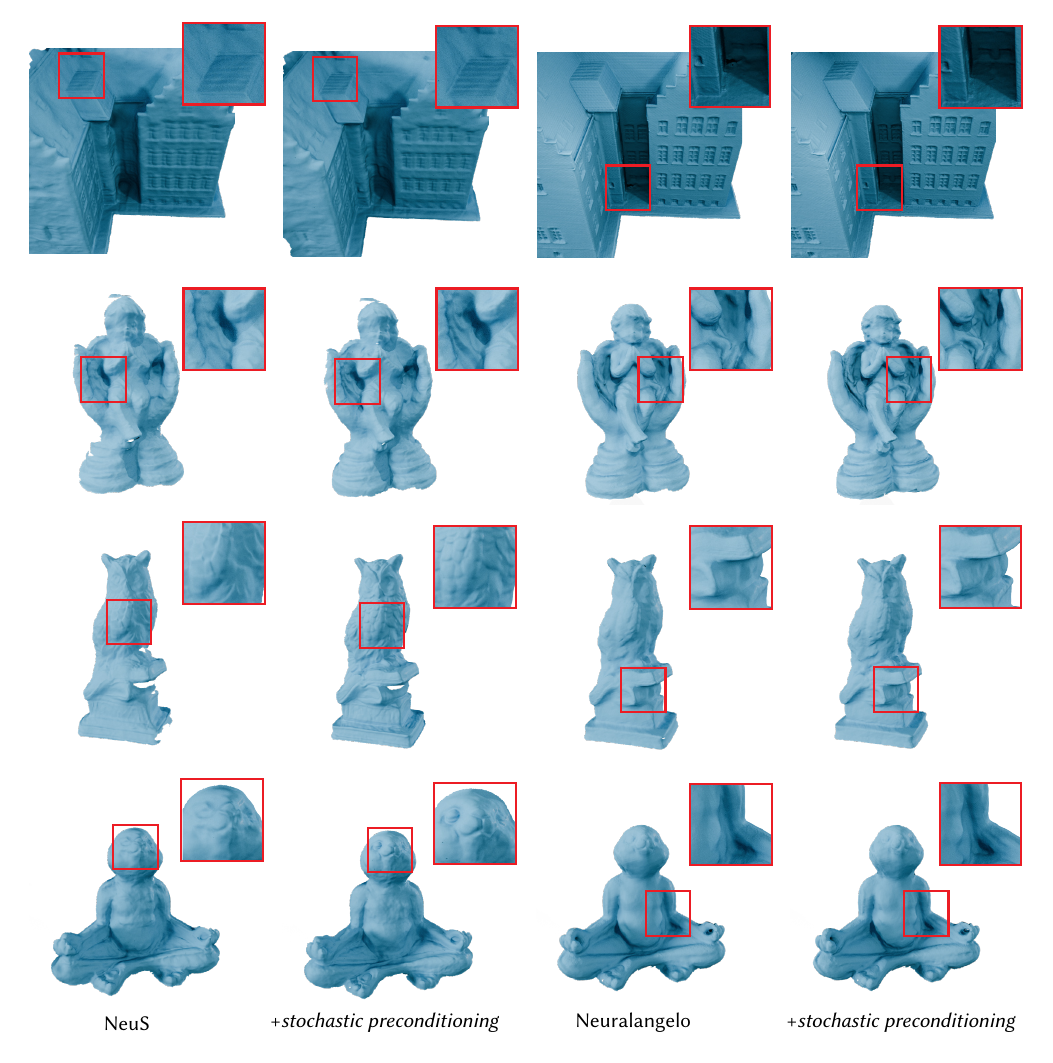}
    \vspace{-6.0mm}
    \caption{
        \label{fig:3d-reconstruct-comprison}
        Here we compare extracted meshes from NeuS~\cite{neus} and Neuralangelo~\cite{neuralangelo}, with and without stochastic preconditioning.
        We find that stochastic preconditioning offers a small but consistent improvement of high-frequency detail in the fitted result, especially in low-visibility regions.
    }
\end{figure*}

We do not suggest any new architectures or tasks, but rather we study how stochastic preconditioning can be applied to a wide range of existing settings, augmenting or even replacing specialized techniques developed for the same purpose (\autoref{fig:teaser}).
Practical benefits include improved output quality across several problems and datasets (Figures~\ref{fig:pn-sdf-comparison}, \ref{fig:few-shot-nerf} and~\ref{fig:3d-reconstruct-comprison}), as well as robustness across a wide range of hyperparameters (\autoref{fig:param-sweep}).
Most importantly, this technique is extremely easy to apply, often just a few lines of code, and is broadly applicable to many formulations---the minor code changes required are shown in \autoref{fig:code-changes}.
For problems in which customized hierarchical strategies already exist, much of their benefit could be had from simply using stochastic preconditioning instead (\autoref{fig:relufield}).
In settings where such techniques have not yet been deployed, there is an immediate and nearly-free boost in quality to be had by enabling stochastic preconditioning (\autoref{fig:3d-reconstruct-comprison}).

\begin{figure}[t]
    \centering
    \begin{minipage}{0.90\linewidth}
\begin{lstlisting}[language=Pseudocode,mathescape=true,escapeinside={@!}{!@}]
def forward():
    x = generate_samples()
    # Apply stochastic preconditioning
    @!\ttfamily{\textcolor{darkgreen}{x += $\alpha$ * torch.randn\_like(x)}}!@
    return f(x)
\end{lstlisting}
    \end{minipage}
    \vspace{-0.5em}
    \caption{
        \label{fig:code-changes}
        Example integration of stochastic preconditioning in an existing implementation of a neural field-based method.
    }
\end{figure}

\section{Related Work}
\label{sec:related-work}

Since a comprehensive overview of neural fields is out of scope, we recommend the course by \citet{Takikawa2023NeuralFieldsCourse}.
Here, we focus on neural field architectures and optimization that are directly relevant.
See \autoref{sec:experiments} for references on downstream task formulations.

\paragraph{Frequency domain.}
Neural field representations quickly moved past naively encoding a field in a MLP.
One line of work leverages frequency-domain encodings of the input coordinates, enriching the input space with high-frequency features~\cite{fourierfeat}.
The use of this technique for radiance fields~\cite{nerf} was key to the broad adoption of neural fields, and continues to be effective in subsequent formulations~\cite{neus}.
SIREN~\cite{siren} builds an architecture entirely from periodic activations to exploit their properties, while BACON~\cite{bacon}, BANF~\cite{banf} and TUNER \cite{tuner} deepen this perspective with explicit control over the frequency spectrum of the signal.
Our stochastic preconditioning likewise leverages frequency-adapted optimization via a blur operation, but in a manner which is compatible with any neural field architecture, and can be applied in conjunction with frequency-space representations for further improvements (\autoref{fig:pn-sdf-comparison-encodings},~\autoref{tab:nsr_small_dtu}).
Another line of work considers signal processing-like operations on neural fields~\cite{xu2022signal,nsampi2023neural} including a related stochastic sampling perspective in~\cite{rebain2024neural}; our work focuses on blurring as an optimization technique, rather than tasks where a filtered signal arises in the desired input or output.

\paragraph{Hybrid feature grids.}
Another approach is to augment the neural field with trainable input features sampled from some backing store or grid~\cite{yu_and_fridovichkeil2021plenoxels}.
Instant~NGP \cite{instantngp} fetches features from a multi-resolution hashgrid encoding, a technique that has since become widely used for large-scale and high-performance neural fields~\cite{neuralangelo,zipnerf,compact}.
Though efficient, naively-applied hash-encoded neural fields seem particularly prone to spurious local minima---adding our stochastic preconditioning offers a significant improvement (Figures~\ref{fig:pn-sdf-comparison}).
Triplane encodings map or learn 3D features to a triplet of planar grids orthogonal to the coordinate axes~\cite{pet}, while \citet{relufield} explore the limit case of a single nonlinearity applied directly to grid features without MLP.
In both settings, our approach again has benefits (\autoref{fig:pn-sdf-comparison-encodings}).

\paragraph{Optimization.}
Rather than modifying the objective or architecture of any given problem, our approach modifies the training process.
Coarse-to-fine approaches have a long history in optimization, as well as specifically in neural fields~\cite{freenerf, neuralgaussian, neuralangelo, bungeenerf, relufield}, and several important downstream applications~\cite{c2fspeech, c2fcontour}.
Neural fields which leverage  hierarchical feature stores~\cite{nglod} take particular advantage of coarse-to-fine optimization by adaptively unlocking levels of the hierarchy ~\cite{neuralangelo,relufield,ringnerf,sape,spatialhash}.
\citet{freenerf} take a similar approach in the frequency-domain~\cite{fourierfeat}, gradually increasing the frequency range of the input encoding.
\citet{datatransformation} permute pixel coordinates on images to accelerate optimization. Another work specifically addresses the shortcomings of neural radiance field training from a gradient scaling perspective~\cite{nerfgradscaling}.
These techniques are simple and effective, but specific to the internals of the representation---stochastic preconditioning offers similar coarse-to-fine optimization with a generic approach applicable to any neural field.

\paragraph{Stochastic perturbations in machine learning}
Perturbation of inputs during neural network training has a rich history in the machine learning literature.
\citet{sietsma1991creating} first proposed adding noise to network inputs to improve generalization, though in simpler settings.
The theoretical foundations were later established by \citet{bishop1995training}, showing that input noise injection is equivalent to Tikhonov regularization for specific loss functions.
\citet{grandvalet1997noise} further connected Gaussian input noise to heat kernel diffusion, providing additional theoretical guarantees.
Adding noise to other quantities during training, such as the activations, was also explored \cite{an1996effects,camuto2020explicit}.
While these classical works focus on perturbations to improve generalization and aim to provide theoretical guarantees in particular settings, our approach instead leverages stochastic smoothing specifically to guide optimization trajectories and avoid local minima in the context of spatial neural fields.

\section{Background}

We start by reviewing concepts and establishing notation for neural fields, gradient preconditioning, and Gaussian smoothing.

\label{sec:background}
\subsection{Neural Fields (and Other Fields Too)}
A neural field encodes a signal on a domain as a function $f_\theta: \mathds{R}^m \rightarrow \mathds{R}^n$, parameterized by some kind of neural network with parameters $\theta$ which maps input coordinates to signal values.
For example, a neural field representing an image would have spatial image locations $(x,y)$ as function inputs with $m=2$, and the RGB color at that location as function outputs with $n=3$; the image is encoded by the parameters of the neural network.

In fact, some of the field representations discussed in this work are not strictly neural, but more generally any parametric function $f_\theta$ used to encode a field. For example, \citet{relufield} sample values from a grid without evaluating any MLP.
Stochastic preconditioning applies more broadly to any notion of a queryable field, although we continue to use the term neural field for familiarity.

\subsection{Gradient Preconditioning}
\label{subsec:gradient-precond}
In numerical linear algebra, preconditioning refers to transforming an ill-conditioned system $A\vx = \vb$ into an equivalent system with better numerical properties by carefully choosing a non-singular matrix $M$ and solving $A M^{-1}\vy = \vb$ followed by $M\vx = \vy$ instead.

\begin{wrapfigure}[6]{R}{0.4\linewidth}
\includegraphics[width=0.2\textwidth,trim=0.5cm 0.0cm 0.0cm 0.5cm]{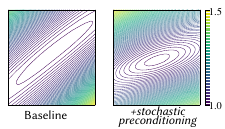}
\label{fig:preconditioning-loss-landscape}
\end{wrapfigure}

In the context of optimization, the intuition for preconditioning is that it changes the geometry of the parameter space so that the level sets of the loss landscape become more isotropic, improving convergence. %
In homage to this perspective, we plot the loss landscape around a minimum of a neural field (inset), following \citeauthor{lossscape}~\shortcite{lossscape}.
The model trained with our stochastic preconditioning technique has a somewhat more isotropic landscape, aligning with the intuition behind preconditioning.

\begin{figure}[t]
  \centering
  \includegraphics[width=\linewidth]{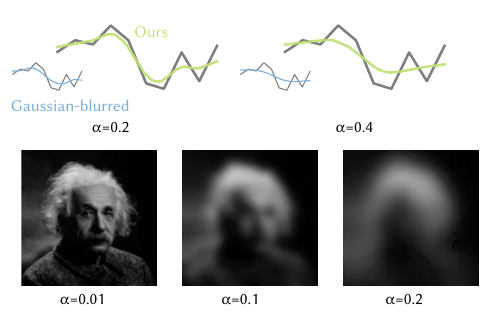}
  \vspace{-5.0mm}
  \caption{
    A simple example of fitting stochastically-blurred neural fields to 1D (\emph{top}) and 2D (\emph{bottom}) signals.
    Sampling from the blurred $f(\vx + \bm{\delta})$ while fitting yields a field $f(\vx)$ which is itself blurred.
    The level of blur closely approximates the corresponding blur of the true signal (\emph{top, inset}).
  }
  \label{fig:stc-effect}
\end{figure}

\subsection{Gaussian Smoothing of Signals}
Gaussian smoothing, or blurring, is a widely-used operation in signal processing, where it can be viewed as damping of high-frequency content or the action of a heat flow.
It can be defined by convolving a signal $f$ with a Gaussian kernel $g$,
\begin{equation}
\label{eq:gaussian-blur}
\textrm{Blur}_\alpha[f]
    = f \ast g
    = \int_\Omega{f(\vx-\bm{\uptau}) \, G_\alpha(\bm{\uptau})d\bm{\uptau}},
\qquad
\end{equation}
where $\Omega$ is the spatial domain, and $G_\alpha(\cdot)$ the Gaussian density.
$\alpha$ gives the isotropic blur width, also known as the standard deviation of the normal distribution.

It is straightforward to evaluate Gaussian smoothing for discrete signals on a grid, but nontrivial to do the same for implicit neural fields.
Past work has addressed this challenge by designing specific band-limited network architectures and training procedures for low-pass filtered neural field signals, motivated by the benefits of enabling level-of-detail post-processing in downstream applications \cite{banf, neuralgaussian, ringnerf, mipnerf, refnerf}.

\subsection{Stochastic Smoothing Evaluation}
We can rewrite \autoref{eq:gaussian-blur} as the following expectation over a random variable $\delta$ drawn from the corresponding normal distribution:
\begin{equation}
\textrm{Blur}[f] = \int_\Omega f(\vx+\bm{\uptau})G(\bm{\uptau}) = \mathds{E}[f(\vx+\bm{\delta})],
\quad
\bm{\delta} \backsim \mathcal{N}(0,\alpha).
\end{equation}
Thus simply perturbing query locations according to a normal distribution gives a stochastic estimate of the blurred neural field.
We will leverage this sampling to access the blurred field $\textrm{Blur}_\alpha[f_\theta]$ in an implicit, online manner during optimization.

\section{Stochastic Preconditioning}
\label{sec:method}

\begin{wrapfigure}[6]{L}{0.3\linewidth}
\includegraphics[width=0.18\textwidth,trim=0.5cm 1cm 0.1cm 0.5cm]{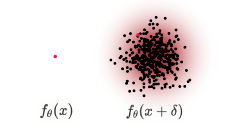}
\label{fig:sampling}
\end{wrapfigure}

We propose to leverage stochastic smoothing as a preconditioner to improve neural field optimization.
Concretely: during training, any time the neural field $f_\theta$ would be queried at position $\vx$, we perturb the sample location according to Gaussian noise to instead query at location $\vx + \bm{\delta}$, with $\bm{\delta} \backsim \mathcal{N}(0,\alpha)$ (see inset).
Otherwise, sampling, loss evaluation, and auto-differentiation are performed as usual for the representation and task at hand.
This can be viewed as a stochastic approximation to optimizing through the blurred field $\textrm{Blur}_\alpha[f](\vx)$.

Two important details complete the method:
(1) we always approximate the blurred field with a single sample, meaning the computation cost is unchanged,
and (2) the scale $\alpha$ is annealed from an initial value down to $0$ during training, such that the final iterations of training are performed on the un-blurred field, and we are left with an ordinary neural field---see \autoref{sec:perturbation_scale} for details and \autoref{sec:Analysis} for analysis.

In implementation, this technique requires changing just a few lines of code (\autoref{fig:code-changes}).
Like past techniques, it effectively manipulates the frequency spectrum of the field, but it does so in a simple, purely implicit manner, which is compatible with any neural field representation, making it straightforward to apply in practice and within existing codebases.

\paragraph{Blurring the field vs. blurring the supervision}
An important subtlety of this approach is that we do not blur the \emph{supervision} (ground-truth data), as has been explored in~\cite{mipnerf,zipnerf,relufield}.
Rather, we blur the \emph{field} being optimized, as a preconditioner.
The primary reason is that blurring the supervision is only possible in directly supervised problems such as fitting to a known signal, whereas blurring the field is fully general, and improves optimization even in regions where the field is governed only by a regularizer (\autoref{sec:neural-sdf}).
Perhaps surprisingly, this more general approach still causes the field $f_\theta$ to be optimized at low-frequencies first (\autoref{fig:stc-effect}).
We hypothesize that this is because stochastic preconditioning damps high-frequency components, and thus also their gradients, during optimization.

\subsection{Boundary Handling}

\begin{figure}[t]
    \centering
    \includegraphics[width=\linewidth]{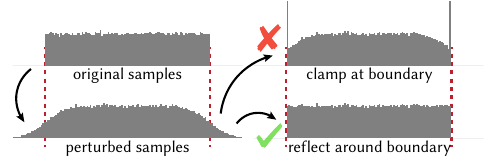}
    \vspace{-6.0mm}
    \caption{
        \textbf{Boundary handling.}
        Stochastic preconditioning will perturb samples outside of bounded domains.
        Clamping would incorrectly concentrate samples on the boundary, while reflecting around the boundary retains the expected uniform distribution.
    }
    \label{fig:boundary}
\end{figure}

Neural fields are often defined only on a bounded domain, such as a normalized $[0,1]^3$ box, but the perturbed query points $\vx + \bm{\delta}$ from stochastic preconditioning can fall outside of that domain.
Naively clamping samples to the boundary creates artifacts, as it results in vastly more samples being drawn along on the boundary during optimization.
Rejecting and re-sampling these points avoids this issue, but complicates implementation.
Instead, \emph{reflecting} samples across the boundary is easily evaluated in constant time, and preserves a uniform sampling distribution, as shown in \autoref{fig:boundary}.
This reflection can be implemented with a coordinate-wise modulo operation, handling even rare multiple-reflections.
For example, perturbed samples $x$ can be reflected in the common $[0, 1]$ domain as
\begin{equation}
x \ \gets \begin{cases}
\ \texttt{mod}(x,2) & \text{if } \quad \texttt{mod}(x,2) \leq 1, \\
\ 2 - \texttt{mod}(x,2) & \text{if } \quad \texttt{mod}(x,2) > 1.
\end{cases}
\end{equation}

\subsection{Choosing the Perturbation Scale $\alpha$}
\label{sec:perturbation_scale}

Stochastic preconditioning requires choosing the blur scale $\alpha \in \mathds{R}^{\geq 0}$, measured in spatial length units.
Large $\alpha$ corresponds to a large low-frequency blur, while $\alpha = 0$ leaves the field unchanged.
In principle $\alpha$ could be a more general covariance matrix, but we find a scalar sufficient in our experiments.
Motivated by coarse-to-fine optimization, we decrease $\alpha$ from an initial $\alpha_0$ to $0$ during the optimization process.
The details of this policy could be adjusted based on the problem setting, but we recommend exponentially decaying alpha from $\alpha_0 = 1-2\%$ of the diagonal length of the bounding-box, to $0$ at 1/3 of training or less, after which stochastic preconditioning is effectively disabled.
This policy was found experimentally, see \autoref{sec:Analysis} for a small study.
The final result is an ordinary neural field, which can be used in downstream tasks without concern for the stochastic preconditioning used during training.

\subsection{Spatially Adaptive $\alpha$}
\label{sec:spatially-adaptive-scaling}

We also consider an optional extension of this method to \emph{spatially-varying} scale factors $\alpha$, storing $\alpha(\vx)$ on a regular grid, which is initialized to $\alpha_0$ and optimized during training as an additional degree of freedom.
Each stochastic-preconditioned query $f_\theta(\vx)$ becomes
\begin{equation}
f_\theta(\vx + \bf{\delta}) \qquad \bf{\delta} \backsim \mathcal{N}(0,\alpha(\vx)),
\end{equation}
and we use the standard ``reparameterization trick'' to autodifferentiate through the task objective with respect to $\alpha(\vx)$, by evaluating $\bf{\delta} \gets \alpha(\vx) \mathcal{N}(0,1)$.

Interestingly, we find that even without any additional supervision or guidance, $\alpha(\vx)$ naturally converges to encode a level of detail for the target field across the domain (\autoref{sec:spatially_varying_results}, \autoref{fig:alpha-map}).
This procedure applies generally without any need for analysis of a known ground-truth signal.
Intuitively, gradients to reduce $\alpha(\vx)$ will arise in regions where the desired signal would have content at spatial frequencies smaller than $\alpha(\vx)$, so $\alpha(\vx)$ is gradually optimized to that frequency, but no smaller.
Furthermore, using this spatially-varying field in place of a constant schedule is a promising alternative in practice, and may even offer small improvements in optimization (see \autoref{tab:nsr_spatial_vs_scalar}).

\section{Experiments}
\label{sec:experiments}

We validate and explore stochastic preconditioning on a range of neural field representations and tasks.
We make a distinction between \emph{indirectly supervised} and \emph{directly supervised} tasks.
In an indirectly supervised task, the goal is to solve a nontrivial inverse problem guided by incomplete supervision; we consider surface reconstruction from images and oriented point clouds, and neural radiance fields (NeRFs).
In directly supervised tasks, the goal is to encode an already-known signal into a neural field; we consider signed distance functions and images.
In general our method is most beneficial in the indirectly supervised setting, as they are difficult optimization problems with many local minima.

In all experiments, we compare stochastic preconditioning to the baseline neural field optimization at equal training iteration counts.
Adding noise is computationally inexpensive, so there is otherwise little change to the runtime or other characteristics of a method, unless \eg a hierarchy is intentionally removed, as in \autoref{sec:exp-relu-fields}.
For full details of experimental configurations and additional results, please see the \textbf{supplemental text and included webpage}.

\subsection{Neural Signed Distance Functions}
\label{sec:neural-sdf}

Signed distance functions (SDFs) represent closed surfaces via an implicit function $f: \mathds{R}^3 \to \mathds{R}$ such that $|f(\vx)|$ is the distance to the surface with $f(\vx) < 0$ inside and $f(\vx) > 0$ outside; they are widely-used in surface processing and reconstruction with neural fields.

\subsubsection{Neural Surface Reconstruction from Oriented Point Clouds}

\begin{figure*}[t]
    \centering
    \vspace{-2.0mm}
    \includegraphics[width=\linewidth]{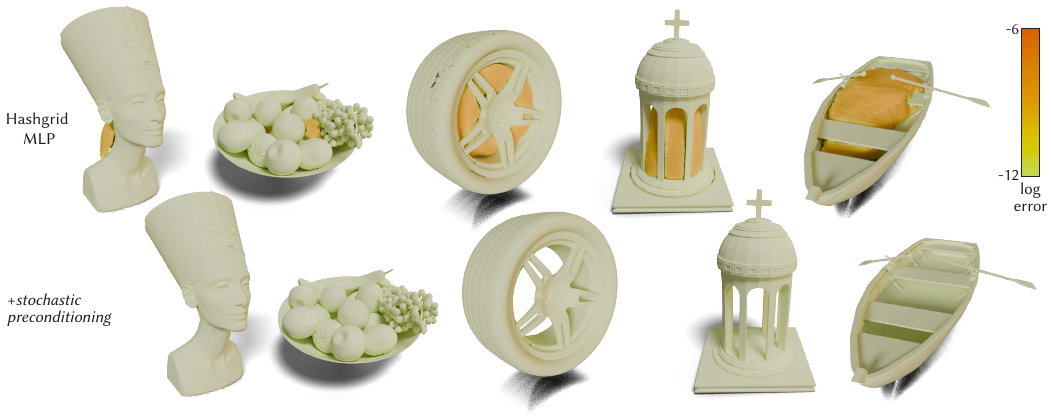}
    \vspace{-8.0mm}
    \caption{
        \textbf{SDFs from oriented point clouds.}
        Stochastic preconditioning reduces artifacts and improves quality when fitting SDFs to point clouds, here using
        an INGP Hashgrid MLP ~\cite{instantngp} as the field representation, along with geometric initialization~\cite{sal}.
        Here we visualize Chamfer error against known ground-truth.
        The standard approach exhibits artifacts which are not resolved by training, while stochastic preconditioning successfully converges to a high-fidelity surface.
    }
    \label{fig:pn-sdf-comparison}
\end{figure*}

Surface reconstruction from point clouds is a classic task in geometry processing; the input is a point cloud with positions and normals $\mathcal{X}=\{\vx_i \in \mathds{R}^3,\bf{n}_i \in \mathds{R}^3\}$ presumed to be sampled from some underlying surface, and the output is that surface.
In the neural fields context, we seek to optimize a scalar field to be the SDF of the reconstructed surface given an input point cloud~\cite{impliciteik, sal, siren}.

As a representative method for this task, we adopt the objective formulation from \textit{SIREN}~\cite{siren}
\begin{equation}
    \label{eq:siren-total-loss}
    \mathcal{L}_{\text{total}} =
                            \lambda_0 \, \mathcal{L}_{\text{surface}}
                          + \lambda_1 \, \mathcal{L}_{\text{normal}}
                          + \lambda_2 \, \mathcal{L}_{\text{eik}}
                          + \lambda_3 \, \mathcal{L}_{\text{offsurface}},
\end{equation}
where $\mathcal{L}_{\text{surface}}$ encourages $f_\theta(\bf{p}_i) \approx 0$ and $\mathcal{L}_{\text{normal}}$ encourages the field gradient at on-surface points to match given normals $\nabla f_\theta(\bf{p}_i) \approx \bf{n_i}$. The regularizer $\mathcal{L}_{\text{eik}}$ is an eikonal term encouraging $f_\theta$ to be a distance function $|\nabla f_\theta(\vx)| = 1$, and $\mathcal{L}_{\text{offsurface}}$ regularizes $f_\theta$ to have large magnitude away from the surface.
The first two terms are evaluated over the input point set, while the regularizers are sampled throughout space.

We consider several field representations, including INGP \cite{instantngp}, Fourier feature fields~\cite{fourierfeat}, neural triplanes~\cite{pet} and sinusoidal INRs \cite{finer} ---we find that stochastic preconditioning improves quality for all of these representations, as shown in \autoref{fig:pn-sdf-comparison-encodings}.
The benefit is greatest with the widely-used fast INGP hashgrids, where it resolves spurious floating artifacts (\autoref{fig:pn-sdf-comparison}).

\citet{sal} propose a popular \textit{geometric initialization} for fitting SDFs, by carefully initializing the solution to a sphere.
This technique is highly effective for objects, but leads to robustness problems for geometry which is far from spherical, such as complex indoor scenes (\autoref{fig:teaser}).
In \autoref{tab:pn-sdf-chamfer-table}, we measure Chamfer distance against ground truth meshes, and find that stochastic preconditioning alleviates the need for this initialization---optimizing a preconditioned field with no special initialization often outperforms the carefully initialized field.

\begin{table}[!t]
    \centering
    \caption{
        \label{tab:pn-sdf-chamfer-table}
        Surface reconstruction from oriented point clouds, measuring mean Chamfer distance against the true surface (lower is better), on several models.
        Stochastic preconditioning improves accuracy even without special initialization strategies.
    }
    \vspace{-0.5em}
    \resizebox{0.99\linewidth}{!}{
        \begin{tabular}{lccccc}
        \toprule
        & Nefertiti & Cow & Bunny & Buddha & Armadillo \\
        \midrule
        \INGP ~\cite{instantngp} & 3.57e-3 & 3.34e-3 & 3.70e-3 & 3.14e-3 & 2.49e-3 \\
        + geom. init. \cite{sal} & 2.82e-3 & 3.69e-3 & 4.24e-3 & 1.44e-3 & 7.25e-4 \\
        + \emph{stochastic preconditioning} & \textbf{7.29e-4} & \textbf{9.20e-4} & \textbf{9.01e-4} & \textbf{8.38e-4} & 7.10e-4  \\
        + geom. init. \& \emph{stochastic preconditioning} & 7.89e-4 & 9.94e-4 & 9.98e-4 & 9.27e-4 & \textbf{6.80e-4}  \\
        \bottomrule
        \end{tabular}
    }
\end{table}

\renewcommand\arraystretch{0.9}
\begin{table}[!t]
    \centering
    \caption{
        \label{tab:nsr_small_dtu}
        Surface reconstruction from images, measuring mean PSNR and Chamfer Distance evaluated across 15 scenes in the DTU dataset~\cite{dtu}.
        Stochastic preconditioning improves quality, especially for hashgrid based representations. See \autoref{supp:tab:nsr_psnr_dtu} and \autoref{supp:tab:nsr_chamfer_dtu} for full tables.
    }
    \begin{tabularx}{\linewidth}{bss}
        \toprule
         & Avg. PSNR & Avg. Chamfer  \\
        \midrule
        NeuS & 27.29 & 1.82  \\
        + \emph{stochastic preconditioning} & 27.51 & \textbf{1.45} \\
        + hashgrid & 18.60 & 4.58 \\
        + hashgrid \& \emph{stochastic preconditioning} & \textbf{28.26} & \textbf{1.45}  \\
        \midrule
        Neuralangelo & 35.87 & 0.87 \\
        + \emph{stochastic preconditioning} & \textbf{36.26} & \textbf{0.76} \\
        \bottomrule
    \end{tabularx}
\end{table}
\renewcommand{\arraystretch}{1.3} %

\subsubsection{Neural Surface Reconstruction from Images.}
\label{sec:neural-surface-reconstruction-from-images}

A more challenging reconstruction task is to recover surfaces from posed images of a scene via differentiable volumetric rendering of an SDF field.
Much work has adapted the NeRF-like~\cite{nerf} formulation~\cite{neus,volsdf,neuralangelo,unisurf}, defining volumetric density as a function of an SDF field, and rendering that density along with auxiliary field outputs for color or directional radiance.
These approaches extract high-fidelity surfaces when successful, but can be challenging to fit due to the difficult inverse nature of the problem.

We consider three baseline methods for this problem:
NeuS~\cite{neus}, a variant of NeuS replacing the frequency encoding with a hashgrid encoding, and Neuralangelo~\cite{neuralangelo}, which includes a specific coarse-to-fine scheme coupling level unlocking with numerical gradient estimation.
For each method, stochastic preconditioning amounts to adding just a few lines of code to the existing implementation.
When evaluating spatial gradients with finite differences, we apply the same noise to all samples in the stencil.

We train each method with and without stochastic preconditioning on 15 scenes from the DTU dataset as in prior work~\cite{dtu}, and compute the average PSNR scores on all the images of the evaluation set as well as the Chamfer distance to the ground truth mesh (\autoref{tab:nsr_small_dtu}).
Stochastic preconditioning improves quality of most scenes when added to all three methods, again with the most dramatic improvements for hashgrid-based representations.
See \autoref{supp:tab:nsr_psnr_dtu} and \autoref{supp:tab:nsr_chamfer_dtu} for full tables in the supplemental document.

\subsection{Neural Radiance Fields}

Neural Radiance Fields (NeRF)~\cite{nerf} represent the appearance of a scene as a neural field encoding the density and outgoing radiance from each point in space, rendered via volumetric sampling along rays.
In \autoref{sec:neural-surface-reconstruction-from-images} we already considered methods which optimize SDFs via NeRF-like rendering; the approaches here use more general density in place of SDFs, which may yield higher visual fidelity but lack an underlying surface representation.

We evaluate primarily on two cases where overfitting artifacts are particularly likely to occur: sparse supervision and ReLU fields~\cite{relufield} without hierarchical training.

\begin{figure}[t]
    \centering
    \includegraphics[width=\linewidth]{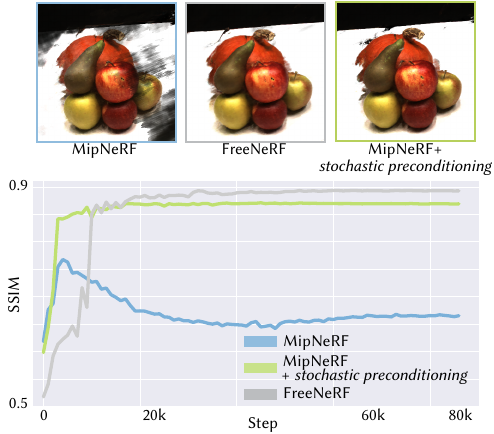}
    \vspace{-6.0mm}
    \caption{
        \textbf{Sparse-view NeRF.}
        We train MipNeRF~\cite{mipnerf}, FreeNeRF~\cite{freenerf} and MipNeRF with our stochastic preconditioning on scene 63 of the DTU dataset~\cite{dtu} with just 6 images as input supervision.
        Nearly achieves the same benefits as FreeNeRF while being compatible with many representations and tasks.
        See the supplement for additional results.
        \label{fig:few-shot-nerf}
    }
\end{figure}

\paragraph{Sparse Supervision}
\label{sec:nerf-sparse-supervision}
First we consider the sparse-view scenario, fitting NeRFs to a comparatively small number of images, which greatly exacerbates the optimization challenges of local minima and overfitting to floater artifacts.
We adopt the setting of FreeNeRF~\cite{freenerf}, a recent approach tackling this challenge via control of frequency-domain encoding and near-camera geometry.
\autoref{fig:few-shot-nerf} shows a comparison against ordinary MipNeRF~\cite{mipnerf} in this setting as a baseline, as as well as simply adding our stochastic preconditioning to MipNeRF.
As expected, MipNeRF struggles in the sparse view scenario, but adding stochastic preconditioning yields high quality results on-par with FreeNeRF---and while FreeNerf is specific to frequency encodings, stochastic preconditioning is applicable to any underlying neural field representation. See \autoref{supp:tab:nerf_few_shots_ssim} in the supplement for additional metrics.

\begin{figure}
    \centering
    \includegraphics[width=\linewidth]{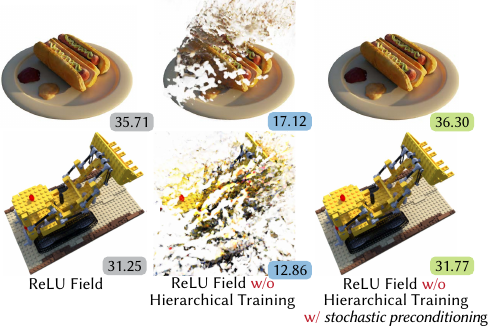}
    \vspace{-6.0mm}
    \caption{
        \label{fig:relufield}
        \textbf{ReLU fields made easy.}
        While the original ReLU fields~\cite{relufield} method performs well on these synthetic scenes (left), it requires a hierarchical training scheme.
        Omitting this scheme, training directly at full resolution, results in catastrophic floater artifacts (center).
        Replacing the coarse-to-fine scheme with our simple stochastic preconditioning method yields on-par or higher-quality results.
        Each result is labeled with the average PSNR.
        See \autoref{supp:fig:relufield_psnr} of the supplemental document for the convergence curve.
    }
    \vspace{-6.0mm}
\end{figure}

\paragraph{ReLU Fields.}
\label{sec:exp-relu-fields}
ReLU fields~\cite{relufield} are a notably \emph{non}-neural representation for NeRFs, exchanging the neural field for a simple trilinearly-interpolated regular grid with a single ReLU nonlinearity after interpolation, akin to prior work optimizing directly on grids~\cite{yu_and_fridovichkeil2021plenoxels}.
To successfully optimize NeRFs, ReLU fields leverage a coarse-to-fine approach, resulting in a four-stage pipeline where each stage uses a finer grid representation and higher-resolution target image; optimization fails catastrophically without this hierarchy.
We find that omitting the hierarchy and simply using stochastic preconditioning instead yields similar or better results (\autoref{fig:relufield}).

In this particular experiment, we remove an existing the coarse-to-fine hierarchy and replace it with stochastic preconditioning---this simplifies the approach, but comes at some cost of increased training time from 3.1 hours to 5.2 hours, since we no longer benefit from fast iterations at the coarse level.

See \autoref{supp:tab:relufield_psnr} in the supplement for additional metrics.

\paragraph{Large-scale reconstruction}
We also perform a preliminary study applying our preconditioning to large-scale photometric reconstruction with Neuralangelo~\cite{neuralangelo} on the Tanks and Temples dataset~\cite{tnt}.
We disabled the coarse-to-fine optimization in Neuralangelo, and instead use stochastic preconditioning, and evaluate PSNR on the foreground rendered image (the background is rendered by a separate background field).
We find that stochastic preconditioning gives a small improvement of $0.16$ PSNR on average to to $32.52$, coming mainly from a significant increase of $+1.16$ on the Barn scene, where artifacts are resolved by preconditioning.
See the Supplement for a breakdown.
Future work will be needed to further investigate effectiveness of this technique for large-scale reconstruction.

\subsection{Analysis}
\label{sec:Analysis}

Here we analyze a few important aspects of our method: the initial alpha value, the choice of noise, and the number of stochastic samples used.

\paragraph{Choice of Initial Alpha}

\setlength{\columnsep}{0.5em}
\setlength{\intextsep}{0em}
\begin{wrapfigure}[6]{R}{0.45\linewidth}
\includegraphics{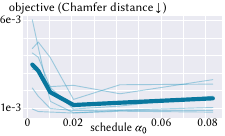}
\label{fig:optimal_alpha}
\end{wrapfigure}
In \autoref{sec:perturbation_scale} we suggest a default policy of scheduling the preconditioning blur scale $\alpha$ from $\alpha_0 = 2\%$ of the domain bounding box diagonal down to $\alpha=0$ during the first $1/3$ of training time.
This policy was found experimentally: the inset figure shows a study in the setting of \autoref{sec:neural-sdf}, each light line corresponds to a different input object, the horizontal axis varies the $\alpha_0$ schedule value, and the vertical axis plots the resulting quality (lower is better). The dark line is the mean, showing an ideal value of around $\alpha_0 = 2\%$ of the bounding-box diagonal.

\paragraph{Choice of Noise}
We default to using Gaussian noise when applying stochastic preconditioning in our experiments for its formal relationship to blur operators, the spectral domain, and heat flow.
We additionally experiment with other noise kernels such as uniform and squared Gaussian noise, for surface reconstruction from points as in \autoref{sec:neural-sdf} and find largely similar results for all kernels, yielding mean Chamfer distance error of 9.1e-4 with a Gaussian kernel, 9.3e-4 with a uniform kernel, and 8.8e-4 with a squared-Gaussian kernel. See the complete set of metrics in  \autoref{supp:tab:pn-sdf-noise-types} in Supplement. 

\paragraph{Single Sample v.s. Multiple Samples}
Our method proposes stochastic preconditioning in expectation, via a single perturbed sample.
However, we also experiment taking additional samples, to more-accurately estimate approximate the blurred field at each evaluation.
In computation, this amounts to averaging multiple samples from $f(\vx + \bm{\delta})$ with independent $\bm{\delta} \backsim \mathcal{N}(0,\alpha)$.
Taking additional samples linearly increases cost, but we do not observe any benefit, likely due to the stochasticity already inherent in neural field optimization.
In the ReLU field setting as in \autoref{sec:exp-relu-fields}, we observe an average PSNR of 33.72 with single sample, 33.74 with two samples, and 33.79 with four samples.
See \autoref{supp:tab:relufield_single_vs_multiple_samples} in the Supplement for full results.

\subsection{Robustness to Hyperparameters.}
\label{sec:robustness-to-hyperparameters}

\begin{figure}[b]
    \centering
    \includegraphics[width=\linewidth]{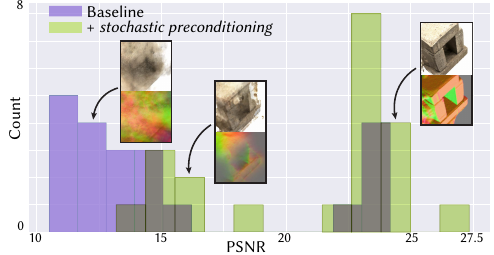}
    \vspace*{-1.5em}
    \caption{
        Stochastic preconditioning increases robustness to hyperparameters, shown here in a histogram of PSNRs from fitting preconditioned and non-preconditioned fields across a range of hyperparameters.
        See~\autoref{sec:robustness-to-hyperparameters} for details.
    }
    \label{fig:param-sweep}
  \end{figure}

Small changes to hyperparameters can easily make the difference between success and getting trapped in a poor local minimum, especially for indirectly supervised tasks.
To explore this effect, we take the NeuS with hashgrid configuration on one scene from \autoref{sec:neural-surface-reconstruction-from-images}, and sweep through several choices of network learning rate, hashgrid learning rate, maximum resolution of the hashgrid, and hash table size.
For each parameter configuration, we fit the field with and without stochastic preconditioning, and measure the final PSNR.
\autoref{fig:param-sweep} shows a clear distribution shift---although even stochastic preconditioning does not succeed for all choices of hyperparameters, it succeeds much more often across a wider range of hyperparameters.
This also provides evidence that the improvement offered by stochastic preconditioning is not an artifact of one particular choice of hyperparameters.

\begin{figure}
    \centering
    \includegraphics[width=\linewidth]{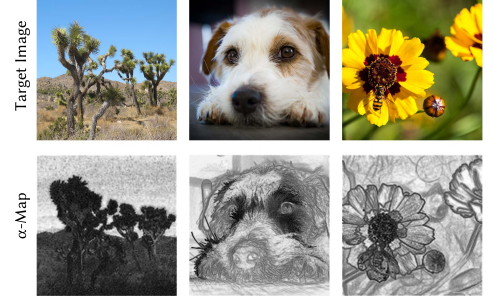}
    \vspace{-6.0mm}
    \caption{
        \label{fig:alpha-map}
        Visualizations of the automatically-optimized spatially-varying $\alpha$-map from fitting target images.
        Photo credits left to right: \copyright Luke Jones \href{https://creativecommons.org/licenses/by/2.0/}{(CC BY 2.0)}, \copyright Jukka Hernetkoski \href{https://creativecommons.org/licenses/by-nc/2.0/}{(CC BY-NC 2.0)}, \copyright Jez \href{https://creativecommons.org/licenses/by-nc-nd/2.0/}{(CC BY-NC-ND 2.0)}.
    }
\end{figure}

\begin{table}[thp]
    \centering
    \vspace*{+0.5em}
    \caption{
        PSNR quality for Neuralangelo~\cite{neuralangelo} reconstruction from images on selected scenes in the DTU dataset, with stochastic preconditioning using spatial-varying scaling \vs a single global scaling.
        \label{tab:nsr_spatial_vs_scalar}
    }
    \begin{tabularx}{\linewidth}{bzzz}
        \toprule
         & DTU 37 & DTU 40 & DTU 55  \\
        \midrule
        Global scheduled $\alpha$ & 29.80 & 34.78 & 31.79 \\
        Optimized spatial-varying $\alpha(\vx)$ & \textbf{30.02} & \textbf{34.80} & \textbf{32.20}\\
        \bottomrule
    \end{tabularx}
\end{table}

\subsection{Spatially-Varying Blur}
\label{sec:spatially_varying_results}

Rather than a global blur radius $\alpha$ decreased with a predetermined schedule, we can instead allow $\alpha(\vx)$ to vary over the domain, optimized automatically according to task objectives (\autoref{sec:spatially-adaptive-scaling}).
In ~\autoref{fig:alpha-map} we apply this technique while fitting neural fields to encode a target image.
The resulting $\alpha(\vx)$ naturally evolves to a good estimation of the frequency distribution of the image contents.
\autoref{tab:nsr_spatial_vs_scalar} shows the result of using the spatially-varying $\alpha(\vx)$ for reconstruction on the DTU dataset with Neuralangelo as in \autoref{sec:neural-surface-reconstruction-from-images}---the automatic, spatially-varying approach matches or even slightly improves quality.

\section{Conclusion}

\paragraph{Limitations and future work.}

Stochastic sampling is an \emph{unbiased} estimator for the blurred field, but sampling a nonlinear loss function with a single sample is a \emph{biased} estimator for the loss on the true blurred field.
We do not observe this to cause problems in practice, but there is opportunity for deeper analysis.
Although stochastic preconditioning improves quality at no additional cost, it misses out on some accelerations offered by hierarchical schemes---optimizing a low-resolution level of a hierarchy is typically very cheap, but applying stochastic preconditioning with a large blur has the same cost as optimizing the full field. 
We only briefly investigate spatially-varying noise distributions (\autoref{sec:spatially_varying_results}); future work may explore additional strategies, including anisotropic distributions, as well as other uses for the optimized $\alpha$ maps

\paragraph{Conclusion.}
We believe that stochastic preconditioning will serve as a valuable tool in the neural fields toolbox, easily deployed to improve performance on a wide variety of problems at minimal cost, especially when custom-designed approaches may be cumbersome.

\section*{Acknowledgments}
We thank Zan Gojcic, Thomas M{\"u}ller, Zian Wang, Sanja Fidler and Alex Keller for their help throughout this work, as well as John Hancock the Dynamic Graphics Project for computing support. 
We are grateful to the artists of 3D models used for demonstrations in this paper, including Turbosquid users Marc Mons, A\_Akhtar, charbavito, Stasma, sougatadhar16, Poly Forge, Vadim Manoli and Digital Fashionwear BD.
Our research is funded in part by NSERC Discovery (RGPIN–2022–04680), the Ontario Early Research Award program, the Canada Research Chairs Program, a Sloan Research Fellowship, the DSI Catalyst Grant program and gifts by Adobe Inc.

\bibliographystyle{ACM-Reference-Format}
\bibliography{references}

\hphantom{a}

\clearpage
\end{document}


\title[\papertitle : Supplemental material]{\papertitle :\\Supplemental material}

\maketitle

\appendix

\vspace*{1em}
\noindent\fbox{%
    \parbox{\linewidth}{%
\textbf{Please see project website for additional results and visualizations: \href{https://research.nvidia.com/labs/toronto-ai/stochastic-preconditioning/}{https://research.nvidia.com/labs/toronto-ai/stochastic-preconditioning/}.}%
}
}
\vspace*{1em}

\section{Experimental details}
\label{sec:experiments-details}

We provide further details on the experiments described in \autoref{main:sec:experiments} of the main paper.

\subsection{Neural Signed Distance Functions}

\subsubsection{Neural Surface Reconstruction from Oriented Point Clouds}

The individual terms of \autoref{main:eq:siren-total-loss} of the main document are as follows:
\begin{align}
    \mathcal{L}_{\text{surface}}
        &= \frac{1}{n} \sum_{i=1}^n \left| f_\theta(\vx_i) \right| \\
    \mathcal{L}_{\text{normal}}
        &= \frac{1}{n} \sum_{i=1}^n (1- \langle \nabla_{\vx_i}(f_\theta),\text{normal}(\vx_i) \rangle) \\
    \mathcal{L}_{\text{eik}}
        &= \int_\Omega \left|\left| |\nabla_{\vx}(f_\theta)| - 1 \right|\right| ~\mathrm{d}\vx \\
    \mathcal{L}_{\text{offsurface}}
        &= \int_\Omega e^{-\alpha|f_\theta(\vx)|}dx.
\end{align}

\paragraph{Hyperparameters.}
For the hashgrid encoding, we use 16 levels with a resolution 16 at the lowest level, a resolution of 2048 at the highest level and hash tables of size $2^{19}$ for each level. 
For the Fourier encoding, we use 6 encoded frequencies. 
For triplane-based feature encoding, we adopt the hyperparameters from \citeauthor{pet}~\shortcite{pet}: resolution $128\times128$ with 24 feature channels, self-attention convolution kernel of window size 8 and 2 heads, and 12 frequencies for the positional encoding. 
For geometric initialization~\cite{sal}, we set the bias term to $0.1$.

For the loss terms coefficients, we used $\lambda_0 = 3 \times 10^3$, $\lambda_1 = 1 \times 10^2$, $\lambda_2 = 5$ and $\lambda_3 = 1 \times 10^2$, following the Siren paper~\cite{siren}. 
We train for 100,000 steps in total with Adam~\cite{adam} with learning rate $1 \times 10^{-3}$ for each experiment.
Note however that 100,000 iterations is much more than needed---we used a large iteration count to ensure that \eg baselines were given the opportunity to recover from initial artifacts.
These same hyperparameters are applied to both the baseline methods and ours.

When enabling stochastic preconditioning, we initialize $\alpha$ to $0.02$, which is 2 percent of the bounding box size, and apply stochastic preconditioning for 2000 steps, after which $\alpha$ is set to zero.

Note that the exponentially-decayed $\alpha$ schedule suggested in the main paper is an equally good option, here we adopted a single-step change for simplicity.

\begin{table}[!t]
    \centering
    \caption{
        \label{tab:pn-sdf-noise-types}
        Surface reconstruction from oriented point clouds with and without stochastic preconditioning using one type of sinusoidal INR called Finer \cite{finer}. Here we measure mean Chamfer distance against the true surface (lower is better) on several models.
    }
    \vspace{-0.5em}
    \resizebox{0.99\linewidth}{!}{
    \begin{tabular}{lccccc}
        \toprule
        
        & Nefertiti & Cow & Bunny & Buddha & Armadillo \\
        \midrule
        FINER & \textbf{4.97e-3} & 9.98e-2  & 7.86e-3 & 8.54e-2 & \textbf{5.28e-3} \\
        + stochastic preconditioning & 5.41e-3 & \textbf{6.45e-3} & \textbf{6.77e-3} & \textbf{6.68e-3} & 6.76e-3 \\
        \bottomrule
    \end{tabular}
}
\end{table}

The results of \autoref{main:fig:pn-sdf-comparison} of the main paper are visualized at equal iteration count for the two methods. However, since each mesh took a different number of training steps to converge, we show extracted mesh at different training steps: \emph{Nefertiti} at 8000 steps, \emph{Fruits bowl} at 10,000 steps, \emph{Tire} at 20,000 steps, \emph{Dome} at 20,000 steps and \emph{Rowboat} at 20,000 steps.

\subsubsection{Neural Surface Reconstruction from Images.}

\paragraph{Baseline methods.}
We compare to NeuS \cite{neus}, which uses a Fourier encoding, a variant of NeuS with hashgrid encoding \cite{instantngp}, and Neuralangelo \cite{neuralangelo}.

\paragraph{Hyperparameters.}
For NeuS, we use 6 frequencies for the positional encoding.
For NeuS with hashgrid encoding, we replace the positional encoding with the multi-resolution hashgrid encoding proposed in \cite{instantngp} within the neural signed distance field.
More specifically, we use 16 levels of hashgrid with resolution of 16 at the lowest level and resolution of 2048 at the highest level and hash tables of size $2^{19}$ and feature size 2 for each level. 
We train for 100,000 steps in total. 
We adopt SDFStudio's~\cite{sdfstudio} re-implementation of NeuS for other training details: \href{https://github.com/autonomousvision/sdfstudio}{https://github.com/autonomousvision/sdfstudio}.

For Neuralangelo, we follow the authors' hyperparameter choice: 16 levels of hashgrid with a resolution 32 at the lowest level and a resolution of 2048 at the highest level, hash tables of size $2^{22}$, and feature size 8 at each level.
We train for 500,000 steps in total.
See the official Neuralangelo implementation for other training details: \href{https://github.com/NVlabs/neuralangelo}{https://github.com/NVlabs/neuralangelo}\\

With stochastic preconditioning, we apply it for 20,000 steps with NeuS-related experiments and 150,000 steps with Neuralangelo.

For NeuS-related experiments, we initialize $\alpha$ to 0.013 and exponentially anneal it to zero during the first 20,000 steps. 
For Neuralangelo, we experimented with both spatially-adaptive scaling and a single global scalar scaling.
For spatially-adaptive scaling, we store the scale factors on a grid of size $128 \times 128 \times 128$ initialized uniformly with value 0.03.
During stochastic preconditioning, we optimize the scaling grid with Adam with learning rate $1 \times 10^{-3}$.
For single global scalar scaling, we set the scale factor $\alpha$ initially to 0.03 and exponentially anneal it to zero during the stochastic preconditioning phase.

\begin{figure}[pt]
    \centering
    \includegraphics[width=\linewidth]{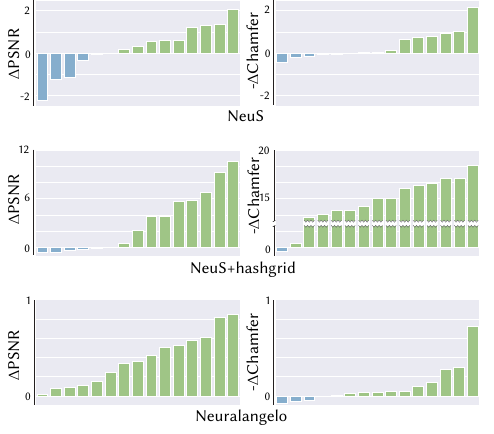}
    \vspace{-6.0mm}
    \caption{
        \label{fig:metrics}
        A sorted bar plot of the difference in PSNR score and negative difference in Chamfer distance between the baselines (NeuS \cite{neus}, NeuS with hashgrid encoding \cite{instantngp}, Neuralangelo \cite{neuralangelo}) and those same baselines augmented by our preconditioning.
        A positive (green) bar indicates that adding stochastic preconditioning improves the metric.
        Each bar represents one scene in the DTU dataset~\cite{dtu}.
        Our method improves quality in most scenes, especially on hashgrid-based representations.
    }
\end{figure}

\paragraph{Quantitative results.}
Detailed quantitative results are shown in \autoref{tab:nsr_psnr_dtu}, \autoref{tab:nsr_chamfer_dtu} and  \autoref{fig:metrics}. 
We also compare spatially-adaptive scaling and global scaling on a small set of scenes and show results in \autoref{main:tab:nsr_spatial_vs_scalar} in the main paper.

\renewcommand\arraystretch{1.3}
\begin{table*}[thp]
    \centering
    \caption{
        Neural surface reconstruction from images.
        Mean PSNR computed over the evaluation set for 15 scenes in DTU dataset~\cite{dtu}.
        \label{tab:nsr_psnr_dtu}
    }
    \resizebox{\linewidth}{!}{
        \begin{tabular}{lccccccccccccccc}
        \toprule
         Methods
         & 24 & 37 & 40 & 55 & 63 & 65 & 69 & 83 & 97 & 105 & 106 & 110 & 114 & 118 & 122  \\
        \midrule
        NeuS & 25.53 & \textbf{23.78} & 25.33 & \textbf{25.25} & \textbf{28.10} & 27.25 & 26.55 & \textbf{26.10} & 26.34 & \textbf{23.79} & 30.67 & 29.12 & \textbf{29.63} & 32.42 & 29.56 \\
        + Stochastic Preconditioning  & \textbf{26.82} & 22.63 & \textbf{25.64} & 25.20 & 25.89 & \textbf{27.85} & \textbf{27.77} & 24.86 & \textbf{26.52} & 23.77 & \textbf{31.24} & \textbf{29.72} & 29.29 & \textbf{33.79} & \textbf{31.64} \\
        NeuS \\
        + Hashgrid Encoding  & \textbf{29.83} & 10.01 & 14.49 & 13.96 & 15.96 & 10.82 & 17.17 & 23.78 & 17.45 & 10.67 & 17.69 & 23.66 & 24.21 & 24.52 & 24.79  \\
        + Stochastic Preconditioning & 28.98 & \textbf{23.15} & \textbf{24.85} & \textbf{27.51} & \textbf{28.45} & \textbf{28.30} & \textbf{27.65} & \textbf{24.80} & \textbf{26.25} & \textbf{25.36} & \textbf{32.38} & \textbf{30.70}& \textbf{30.55} & \textbf{32.40} & \textbf{32.60}  \\
        \midrule
        Neuralangelo & 35.17 & 29.91 & 34.54 & 31.67 & 40.15 & 36.62 & 36.37 & 36.31 & 32.11 & 36.67 & 37.59 & 35.87 & 35.71 & 39.48 & 39.85 \\
        + Stochastic Preconditioning & \textbf{35.42} & \textbf{30.02} & \textbf{34.80} & \textbf{32.20} & \textbf{40.30} & \textbf{37.13} & \textbf{37.22} & \textbf{36.89} & \textbf{32.13} & \textbf{37.01} & \textbf{37.67} & \textbf{36.29} & \textbf{36.07} & \textbf{40.09} & \textbf{40.67} \\
        \bottomrule
        \end{tabular}
    }
\end{table*}

\begin{table*}[thp]
    \centering
    \caption{
        Neural surface reconstruction from images.
        Mean Chamfer distance to the ground-truth mesh on 15 scenes of the DTU dataset~\cite{dtu}.
        \label{tab:nsr_chamfer_dtu}
    }
    \resizebox{\linewidth}{!}{
        \begin{tabular}{lccccccccccccccc}
        \toprule
            Methods
            & 24 & 37 & 40 & 55 & 63 & 65 & 69 & 83 & 97 & 105 & 106 & 110 & 114 & 118 & 122  \\
        \midrule
        NeuS & \textbf{1.428} & \textbf{2.540} & \textbf{1.743} & \textbf{0.448} & 1.408 & \textbf{1.152} & 1.937 & 5.129 & 2.503 & 1.613 & \textbf{0.694} & 2.362 & 0.453 & 2.349 & 1.501 \\
        + Stochastic Preconditioning  & 1.451 & 2.975 & 1.923 & 0.492 & \textbf{1.367} & 1.294 & \textbf{1.205} & \textbf{2.976} & \textbf{2.390} & \textbf{0.690} & 0.753 & \textbf{1.360} & \textbf{0.404} & \textbf{1.722} & \textbf{0.735} \\
        NeuS \\
        + Hashgrid Encoding  & 0.553 & 7.902 & 6.971 & 7.224 & 3.334 & 11.722 & 10.543 & \textbf{4.179} & \textbf{1.975} & 5.346 & 4.381 & 2.072 & \textbf{0.358} & \textbf{1.481} & \textbf{0.692}  \\
        + Stochastic Preconditioning & \textbf{0.526} & \textbf{2.18} & \textbf{1.144} & \textbf{0.393} & \textbf{1.153} & \textbf{1.104} & \textbf{1.265} & 4.698 & 2.323 & \textbf{1.522} & \textbf{0.549} & \textbf{1.592} & 0.406 & 2.051 & 0.862 \\
        \midrule
        Neuralangelo & 0.437 & 0.771 & 0.337 & \textbf{0.326} & 0.854 & 0.784  & 1.760 & 1.374 & 1.957 & \textbf{0.717} & 0.468 & \textbf{1.229} & \textbf{0.328} & 0.812 & 0.830  \\
        + Stochastic Preconditioning & \textbf{0.385} & \textbf{0.721} & \textbf{0.310} & {0.332} & \textbf{0.843} & \textbf{0.683} & \textbf{1.038} & \textbf{1.336} & \textbf{1.684} & 0.795 & \textbf{0.428} & 1.283 & 0.372 & \textbf{0.513} & \textbf{0.683} \\
        \bottomrule
        \end{tabular}
    }
\end{table*}

\begin{table}[thp]
    \centering
    \caption{
        Neural surface reconstruction from images.
        Mean PSNR computed over the foreground rendering on the evaluation set for 6 scenes in Tanks and Temples dataset \cite{tnt}.
        \label{tab:nsr_psnr_tnt}
    }
    \resizebox{\linewidth}{!}{
        \begin{tabular}{lccccccc}
        \toprule
         Methods
         & Barn & Caterpillar & Ignatius & Meetingroom & Truck & Courthouse & Average  \\
        \midrule
        Neuralangelo & 29.52 & \textbf{35.85} & 40.26 & \textbf{28.85} & \textbf{34.37} & 25.29 & 32.36 \\
        + Stochastic Preconditioning & \textbf{30.68} & 35.81 & \textbf{40.30} & 28.69 & 34.27 & \textbf{25.35} & \textbf{32.52} \\
        \bottomrule
        \end{tabular}
    }
\end{table}

\subsubsection{Fully-supervised SDF Fitting}
\label{subsec:fullsdf}

For fully-supervised SDF fitting, we follow the experiment design proposed by \citeauthor{instantngp}~\shortcite{instantngp}, including using the mean absolute percentage error (MAPE) loss:
$$
\mathcal{L}(f_\theta(\vx), d) = \frac{|f_\theta(\vx)-d|}{|d| + \epsilon}, \epsilon=1e^{-2}.
$$
At each iteration, we minimize the mean absolute percentage error (MAPE) on a batch of sample points ($50\%$ on the surface, $25\%$ near the surface, and $25\%$ uniformly within the domain).

\begin{figure}[pt]
    \centering
    \includegraphics[width=\linewidth]{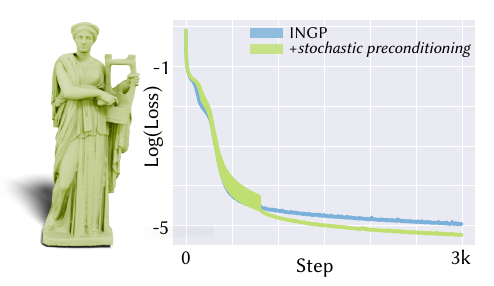}
    \vspace{-6.0mm}
    \caption{
        \label{fig:fullsdf-loss}
        \textbf{Direct SDF fitting.}
        We reproduce the SDF fitting experiment of \citeauthor{instantngp}~\shortcite{instantngp} with direct supervision from a known SDF.
        Stochastic preconditioning enables reaching a significantly lower loss at convergence.
        Please see \autoref{subsec:fullsdf} in main paper for a discussion of the final extracted mesh quality.
    }
\end{figure}

\paragraph{Baseline method.}
We follow the experiment design and architecture of \citeauthor{instantngp}~\shortcite{instantngp}.

\paragraph{Quantitative results.}
In this experiment, adding our stochastic preconditioning allows consistently reaching a much lower loss at convergence, as illustrated on one example in \autoref{fig:fullsdf-loss}.

However, we find that the value of the loss function on its own does not tell the full story.
Indeed, it does not account for non-uniformity in the norm of the field (\ie imperfect SDFs), which is not supervised in the objective function for this setting.

\paragraph{Surface extraction for imperfect SDFs.}
The varying field norm results in a lower loss around the surface, but also in staircase-like artifacts during surface extraction.
We resolve these artifacts by augmenting the Marching Cubes algorithm~\cite{marchingcubes} with a bisection root finding step, instead of linear interpolation, to identify zero-crossing locations.
The inset figure shows the improvements in mesh extraction as the number of bisection steps increases.\\
\makebox[\linewidth]{
    \includegraphics[width=0.75\linewidth]{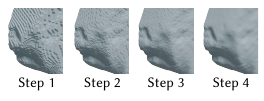}
}

Interestingly, while SP does yield a significantly lower training loss value at convergence, we do not see substantial improvements in the final extracted mesh quality.
We believe this calls for a re-examination of this widely-used objective function, for example incorporating an Eikonal constraint.
This is, however, outside the scope of this paper.

\subsection{Neural Radiance Field}

\subsubsection{Optimization with Sparse Supervision}
\paragraph{Baseline methods.}
We experiment with 6 input views and use the official FreeNeRF \cite{freenerf} codebase, implemented in JAX, to compare with both MipNeRF \cite{mipnerf} and FreeNeRF on the Blender dataset.\\

\paragraph{Hyperparameters}
Both MipNeRF and FreeNeRF use positional encoding with 16 maximum input frequencies.
We follow the implementation in FreeNeRF and concatenate the original coordinates with the positional encodings as input to the network.
All models are trained with the Adam optimizer for a total of 87,891 steps (matching the FreeNeRF default) with a learning rate decaying exponentially from $1 \times 10^{-3}$ to $2 \times 10^{-5}$
and 512 warm-up steps with a multiplier of 0.01.
For FreeNeRF, the frequency regularizer is applied during the first $70\%$ of the training iterations.

We apply stochastic preconditioning to MipNeRF with $\alpha$ initialized to 0.04 and exponentially annealed to zero by iteration 2500.

Numerical results for the experiment of \autoref{main:sec:nerf-sparse-supervision} and \autoref{main:fig:few-shot-nerf} in the main paper are given in \autoref{tab:nerf_few_shots_ssim}. 
See also the supplemental website for additional qualitative results.

\renewcommand\arraystretch{1.0}
\begin{table}[thp]
    \centering
    \caption{
        \label{tab:nerf_few_shots_ssim}
        Neural Radiance Field training with sparse supervision.
        Stochastic preconditioning improves upon the baseline but generally underperforms FreeNeRF~\cite{freenerf}.
    }
    \begin{tabularx}{\linewidth}{bvvvvv}
        \toprule
         Methods &
         21 & 30 & 55 & 63 & 103   \\
        \midrule
        MipNeRF & 0.448 & 0.632 & 0.371 & 0.665 & 0.703 \\
        MipNeRF + Stochastic Preconditioning & 0.582 & \textbf{0.872} & 0.699 & 0.869 & 0.758 \\
        FreeNeRF & \textbf{0.698} & 0.856 & \textbf{0.746} & \textbf{0.893} & \textbf{0.812} \\
        \bottomrule
    \end{tabularx}
    \vspace{1em}
\end{table}

\subsubsection{Optimization with ReLU Fields}
\paragraph{Baseline methods.}
We use the official ReLU Field~\cite{relufield} implementation:\\
\makebox[\linewidth]{
    \href{https://github.com/akanimax/thr3ed\_atom}{https://github.com/akanimax/thr3ed\_atom}
}

The first baseline is the unmodified ReLU field method, which contains a four-stage hierarchical training schedule for coarse-to-fine optimization.
In the second baseline, we remove this hierarchical training schedule.

\paragraph{Hyperparameters.}
We follow the hyperparameters used in the original ReLU field experiment. 
Specifically, we use a dense grid of resolution $256 \times 256 \times 256$ and optimize with the Adam optimizer with a learning rate of $0.03$. 
The original ReLU field baseline employs a hierarchical training curriculum in which the grid is initially optimized at a resolution reduced by a factor of $2^4$. 
The grid is trained for 7,000 iterations per-stage, then is upsampled by a factor of two after each stage.
The grid features is trilinearly-interpolated to initialize the grid for next stage. 
Note that the learning rate exponentially decreases per stage after 3,000 steps.

The second baseline, where this hierarchical curriculum was removed, directly optimizes at full resolution.

We apply stochastic preconditioning for the first 8,000 steps with $\alpha$ initialized to 0.003 and exponentially annealed to zero. 
We also start exponential learning rate decay after 12,000 steps instead of 3,000 steps.

\paragraph{Quantitative results.}
Detailed numerical results for the experiment of \autoref{main:sec:exp-relu-fields} and \autoref{main:fig:relufield} of the main paper are shown in \autoref{tab:relufield_psnr}.

\begin{figure}[pt]
    \centering
    \includegraphics[width=\linewidth]{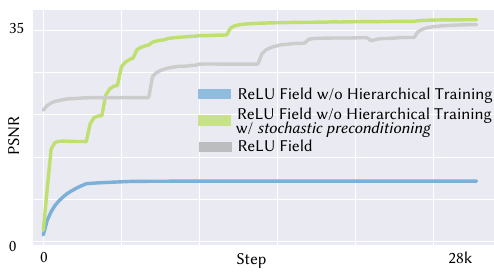}
    \vspace{-6.0mm}
    \caption{
        \label{fig:relufield_psnr}
        \textbf{ReLU Field convergence.}
        We show the PSNR score averaged across test set during training for the ReLU field experiments on the hotdog scene in Blender dataset. Without hierarchical training, ReLU field gets stuck with catastrophic overfitting. Both stochastic conditioning and the original hierarchical training scheme achieves high PSNR scores while ours is much simpler and converges with less training steps without a staged schedule.
    }
\end{figure}

\renewcommand\arraystretch{1.3}
\begin{table}[t]
    \centering
    \caption{
        \label{tab:relufield_single_vs_multiple_samples}
        Average PSNR evaluated across test set with stochastic preconditioning using single samples v.s more.
    }
    \begin{tabularx}{\linewidth}{bzzzz}
        \toprule
         Methods
         & Lego & Chair & Mic & Hotdog  \\
        \midrule
        w/ 1 sample & 33.78 & 31.77 & 33.03 & 36.30 \\
        w/ 2 samples & 33.92 & 31.79 & 32.99 & 36.27 \\
        w/ 4 samples & 34.01 & 31.69 & 33.14 & 36.30 \\
        \bottomrule
    \end{tabularx}
    \vspace{1em}
\end{table}

\begin{table}[!t]
    \centering
    \caption{
        \label{tab:pn-sdf-noise-types}
        Surface reconstruction from oriented point clouds using stochastic preconditioning with different types of noise: gaussian, uniform, and squared gaussian. Here we measure mean Chamfer distance against the true surface (lower is better) on several models.
    }
    \vspace{-0.5em}
    \resizebox{0.99\linewidth}{!}{
        \begin{tabular}{lccccc}
        \toprule
        
        & Nefertiti & Cow & Bunny & Buddha & Armadillo \\
        \midrule
        Gaussian & 7.89e-4 & 9.94e-4 & 9.98e-4 & 9.27e-4 & 6.80e-4 \\
        Uniform & 7.24e-4 & 8.56e-4 & 8.23e-4 & 9.03e-4 & 6.83e-4 \\
        Squared Gaussian & 7.30e-4 & 8.33e-4 & 7.05e-4 & 8.21e-4 & 6.63e-4 \\
        \bottomrule
        \end{tabular}
    }
\end{table}

\renewcommand\arraystretch{1.0}
\begin{table}[thp]
    \centering
    \caption{
        \label{tab:relufield_psnr}
        ReLU field training for novel view synthesis.
        Average LPIPS evaluated across the test set.
        Stochastic preconditioning achieves LPIPS scores comparable with the staged hierarchical training of \citeauthor{relufield}~\shortcite{relufield}, with a simple unified approach.
    }
    \begin{tabularx}{\linewidth}{bvvvvv}
        \toprule
         Methods
         & Lego & Hotdog & Chair & Mic & Ship  \\
        \midrule
        ReLU Field & 0.035 & 0.032 & 0.033 & 0.025 & 0.119 \\
        ReLU Field w/o Hierarchical Training & 0.432 & 0.351 & 0.420 & 0.388 & 0.482 \\
        ReLU Field w/o Hierarchical Training, w/ Stochastic Preconditioning & \textbf{0.031} & \textbf{0.027} & \textbf{0.030} & \textbf{0.023} & \textbf{0.113}\\
        \bottomrule
    \end{tabularx}
    \vspace{1em}
\end{table}

\subsection{Image Fitting}

\begin{table}[t]
    \centering
    \caption{
        \label{tab:psnr-image-alpha-map}
        PSNR score evaluated on a set of five images with baseline method \INGP and our $\alpha$-map-guided adaptive sampling and unlocking as described.
    }
    \resizebox{0.99\linewidth}{!}{
        \begin{tabular}{lccccc}
        \toprule
        Methods
        & Tree & Dog & Flower & Building & Tokyo \\
        \midrule
        \INGP ~\cite{instantngp} & 34.89 & 39.21 & 42.06 & \textbf{40.75} & 32.20 \\
        + $\alpha$-Map-Guided Adaptive Techniques & \textbf{35.23} & \textbf{39.83} & \textbf{42.54} & 40.74 & \textbf{32.47}  \\
        \end{tabular}
    }
\end{table}

\paragraph{Baseline methods.}
For image fitting, we reproduced the experiment from Instant~NGP~\cite{instantngp} using PyTorch and the tiny-cuda-nn library \cite{tinycudann}, and use it as our baseline.

\paragraph{Hyperparameters}
For the baseline with hashgrid encoding, we use 16 levels with a resolution of 16 at the lowest level and a resolution of 2048 at the highest level and hash tables of size $2^{19}$ and feature size 2 for each level. 
We train for 30,000 steps in total using Adam optimizer with learning rate $1 \times 10^{-3}$ that is decayed by 0.6 at 20,000 step.

We apply stochastic preconditioning for the first 6,000 steps with spatially-adaptive scaling to acquire the $\alpha$-maps. 
Specifically, we store the scaling factor in a grid that is the original image's size reduced by a factor of 16. 
We optimize with Adam with learning rate 0.01.

\begin{figure*}[thp]
    \centering
    \vspace{1em}
    \hspace{1.7cm}
    Tree \hfill
    Dog \hfill
    Flower \hfill
    Tokyo \hfill
    Building
    \hspace{1cm}
    \hphantom{a}
    \includegraphics[width=\linewidth]{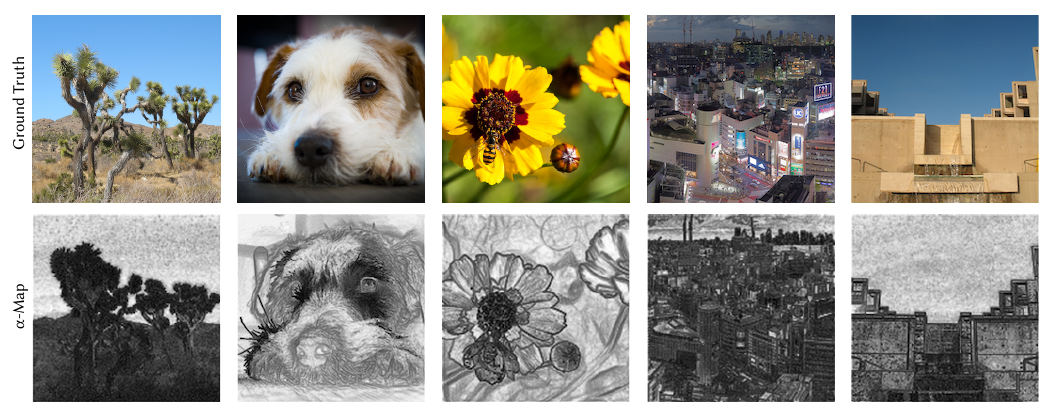}
    \vspace{-6.0mm}
    \caption{
        Image fitting with a spatially-varying $\alpha$ value.
        Photo credit from left to right: \copyright Luke Jones \href{https://creativecommons.org/licenses/by/2.0/}{(CC BY 2.0)}, \copyright Jukka Hernetkoski \href{https://creativecommons.org/licenses/by-nc/2.0/}{(CC BY-NC 2.0)}, \copyright Jez \href{https://creativecommons.org/licenses/by-nc-nd/2.0/}{(CC BY-NC-ND 2.0)}, photograph \copyright zymurgeist \href{https://creativecommons.org/licenses/by-nc-nd/2.0/}{(CC BY-NC-ND 2.0)}, photograph \copyright Trevor Dobson \href{https://creativecommons.org/licenses/by-nc-nd/2.0/}{(CC BY-NC-ND 2.0)}.
    }
  \label{fig:alpha_map_full}
\end{figure*}

Once we acquired the $\alpha$-maps as shown in \autoref{fig:alpha_map_full}, we use the map for a set of spatially-adaptive training techniques based on the intuition that lower $\alpha$ value corresponds to regions with higher frequency details.
We use it for importance sampling during training. 
In addition, we use it for adaptive hashgrid feature masking such that the number of levels of the hashgrid feature used at a query position is inversely related to the magnitude of the optimized $\alpha$ value.

\paragraph{Quantitative results.}
Applying the above $\alpha$-map-based adaptive scheme on a set of five example images, we achieve slightly better PSNR scores on most of them as shown in \autoref{tab:psnr-image-alpha-map}. We show the images mentioned in \autoref{tab:psnr-image-alpha-map} of the main paper, together with their optimized spatially-varying $\alpha$-maps in \autoref{fig:alpha_map_full}.

\vfill  %
\bibliographystyle{ACM-Reference-Format}
\bibliography{references}